\documentclass[letterpaper]{article}
\usepackage[final]{neurips_2019}
\usepackage{algpseudocode, cleveref}
\usepackage{algorithm,comment,amsfonts,amssymb}
\usepackage{algpseudocode, multicol} 
\usepackage{amsmath,soul,bm}
\DeclareMathOperator*{\argmax}{arg\,max}

\usepackage{bbm} 
\usepackage{xcolor}  
\usepackage{graphicx}  
\usepackage{amsthm} 
\usepackage{setspace}
\usepackage{algorithmicx}
\usepackage{bbm}
\usepackage[numbers]{natbib}

\newtheorem{prop}{Proposition}
\newtheorem{lemma}{Lemma}
\newtheorem{defn}{Definition}[section]

\def\aA{a}
\def\aD{d}
\def\A{\mathcal{A}}
\def\D{\mathcal{D}}
\def\Es{\mathcal{E}}
\def\G{\mathcal{G}}
\def\H{\mathcal{I}}
\def\N{\mathcal{N}}
\def\R{\mathcal{R}}
\def\S{\mathcal{S}}
\def\T{\mathcal{T}}

\def\E{\mathbb{E}}
\def\Rb{\mathbb{R}}
\def\W{\mathcal{W}}
\def\Y{\mathcal{Y}}
\def\biggiven{{\,\Big|\,}}

\def\hD{h_{\bar\pi^D\!,\,\mu}}
\def\WD{W_{\bar\pi^D\!,\,\mu}}

\newcommand*{\mybox}[1]{\framebox{#1}}

\usepackage{times}

\title{Non-Cooperative Inverse Reinforcement Learning}

\author{{Xiangyuan Zhang}
  \qquad\quad Kaiqing Zhang  \qquad\quad Erik Miehling \qquad\quad Tamer Ba\c{s}ar\\\\Coordinated Science Laboratory\\ University of Illinois at Urbana-Champaign\\\texttt{\small\{xz7,kzhang66,miehling,basar1\}@illinois.edu}
}

\begin{document} 

\maketitle

\begin{abstract}
Making decisions in the presence of a strategic opponent requires one to take into account the opponent's ability to actively mask its intended objective. To describe such strategic situations, we introduce the non-cooperative inverse reinforcement learning (N-CIRL) formalism. The N-CIRL formalism consists of two agents with completely misaligned objectives, where only one of the agents knows the true objective function. Formally, we model the N-CIRL formalism as a zero-sum Markov game with one-sided incomplete information. Through interacting with the more informed player, the less informed player attempts to both infer, and act according to,  the true objective function. As a result of the one-sided incomplete information, the multi-stage game can be decomposed into a sequence of single-stage games expressed by a recursive formula. Solving this recursive formula yields the value of the N-CIRL game and the more informed player's equilibrium strategy. Another recursive formula, constructed by forming an auxiliary game, termed the dual game, yields the less informed player's strategy. Building upon these two recursive formulas, we develop a computationally tractable algorithm to approximately solve for the equilibrium strategies. Finally, we demonstrate the benefits of our N-CIRL formalism over the existing multi-agent IRL formalism via extensive numerical simulation in a novel cyber security setting. 
\end{abstract}

\section{Introduction}\label{sec:introduction}
{In any decision-making problem, the  decision-maker's goal is characterized by some underlying, potentially unknown, objective function.} In machine learning, ensuring that {the learning} agent does what {the human} intends it to do requires specification of a \emph{correct} objective function, a problem known as \emph{value alignment} \cite{wiener1960some,russell2015research}. Solving this problem is nontrivial even in the simplest (single-agent) reinforcement learning  (RL) settings \cite{arnold2017value} and becomes even more challenging in multi-agent environments \cite{lanctot2017unified,zhang2018fully,zhang2018finite}. Failing to specify a correct objective function can lead to unexpected and potentially dangerous behavior \cite{russell2016artificial}. 

Instead of specifying the correct objective function, a growing body of research has taken an alternative perspective of letting the agent learn the intended task from observed behavior of other agents and/or human experts, a concept known as \emph{inverse reinforcement learning} (IRL)  \citep{ng2000algorithms}. IRL describes a setting where one agent is attempting to learn the true reward function by observing trajectories of sample behavior of another agent, termed \emph{demonstrations}, under the assumption that the  observed agent is acting (optimally) according to the true reward function. More recently, the concept of \emph{cooperative inverse reinforcement learning} (CIRL) was introduced in  \cite{hadfield2016cooperative}. Instead of passively learning from demonstrations (as is the case in IRL), CIRL allows for agents to \emph{interact} during the learning process, which results in improved performance over IRL. CIRL inherently assumes that the two agents are on the same team, that is, the expert is actively trying to help the agent learn {and achieve some common goal.}  However, in many practical multi-agent decision{-making} problems, the objectives of the agents may be misaligned, and in some settings completely opposed \cite{littman1994markov,lin2018multiagent,srinivasan2018actor,zhang2019policy} (\emph{e.g.}, zero-sum interactions in cyber security \cite{ammann2002scalable,miehling2018pomdp}). In such settings, the expert is still trying to achieve {its objective,} but may act in a way so as to make the agent think it has a different objective. Development of a non-cooperative analogue of CIRL to describe learning in these settings has not yet been investigated. 

{In this paper, we introduce the \emph{non-cooperative inverse reinforcement learning} (N-CIRL) formalism. The N-CIRL formalism consists of two agents with completely misaligned objectives, where only one agent knows the true reward function. The problem is modeled as a zero-sum Markov game with one-sided incomplete information. In particular, {at any stage of the game} the information available to the player that does not know the true reward function, termed the \emph{less informed} player, is contained within the information available to the player that knows the reward function, termed the \emph{more informed} player. This {one-sided} information structure allows for a simplification in the form of the beliefs (compared to the fully general asymmetric information case \citep{hansen2004dynamic}) and, more importantly, allows one to define strategies as the stage-wise solutions to two recursive equations \cite{de1996repeated,rosenberg1998duality}. By taking advantage of the structure of these recursive equations, and their corresponding solutions (fixed points), computationally tractable algorithms for approximately computing the players' strategies {are} developed.  

Our primary motivation for developing the N-CIRL formalism is cyber security. In particular, we are interested in settings where the attacker has some intent that is unknown to the defender. In reality, the motivation of attackers can vary significantly. For example, if the attacker is financially motivated, its goal may be to find personal payment details, whereas if the attacker is aiming to disrupt the normal operation of a system, reaching a computer responsible for controlling physical components may be of more interest. This variability in what the attacker values is captured by the N-CIRL formalism through an \emph{intent} parameter that serves to parameterize the attacker's true reward function. The defender, who does not know the true intent, then faces the problem of learning the attacker's intent while simultaneously defending the network. The attacker, knowing this, aims to reach its goal but {may behave} in a way that makes its true intent as unclear as possible.\footnote{An interesting real-world example of such behavior was Operation Fortitude in World War II \citep{hendricks2006feints}.}

Throughout the remainder of the paper, we will refer to the more informed player as the \emph{attacker} and the less informed player as the \emph{defender}. While we use the cyber security example throughout the paper, this is primarily for ease of exposition. The results presented here apply to any zero-sum Markov game setting where one player does not know the true reward. 

{\bf Limitations of Existing {IRL} Approaches.} The application of N-CIRL to cyber security is especially fitting due to the challenges associated with collecting useful attack data. Obtaining accurate attack logs is a computationally formidable task \cite{ji2017rain}. Furthermore, learning from sample equilibrium behavior, as is done in the \emph{multi-agent inverse reinforcement learning} (MA-IRL) settings of  \cite{lin2018multiagent,wang2018competitive,lin2019multi}, is only useful if the goal(s) do not change between learning and execution/deployment. Such an assumption is not appropriate in cyber security settings -- the attacker's goal as well as the overall system structure, may frequently change. This non-stationary behavior necessitates the ability to intertwine learning and execution. N-CIRL provides a formalism for specifying actions that adapt to the information revealed during the game. We illustrate this adaptivity via an illustrative example in Sec. \ref{ssec:example} and {extensive} numerical results in Sec. \ref{sec:exp}.

{\bf Contribution.} The contribution of the present work is three-fold: {\bf 1)} We propose a new formalism for IRL, termed N-CIRL, that describes how two competing agents make strategic decisions when only one of the agents possesses knowledge of the true reward; {\bf 2)} By recognizing that N-CIRL is a zero-sum Markov game with one-sided incomplete information, we leverage the recursive structure of the game to develop a computationally tractable algorithm, termed \emph{non-cooperative point-based value iteration} (NC-PBVI), for computing both players' strategies; {\bf 3)} We demonstrate in a novel cyber security model that the adaptive strategies obtained from N-CIRL outperform strategies obtained from existing multi-agent IRL techniques.

\section{Related Work}\label{sec:related_works}

{Decision-making when the agents are uncertain of the true objective function(s) has been extensively studied within the fields of both RL and game theory.} One standard and popular way to {infer the actual reward function} {is via inverse RL,} {the idea of which} was first  {introduced} by \cite{kalman1964linear} {under the title} of inverse optimal control. Later, \cite{ng2000algorithms} introduced the notion of IRL {with the goal of inferring} the
reward function being optimized by observing the behavior {of} an actor, {termed an \emph{expert}}, over time \citep{ng2000algorithms,abbeel2004apprenticeship,ratliff2006maximum,finn2016guided}. Fundamental to the IRL setting is the assumption that the agent inferring the reward {\emph{passively}} observes the expert's behavior,  {while} the expert behaves {optimally in its own interest} without {knowing} that the agent will later use the observed behavior to learn.

As pointed out in \cite{hadfield2016cooperative}, such an assumption is not valid in certain cooperative settings where the agent and the expert are able to interact in order to achieve some common objective. In fact, IRL-type solutions were shown to be {suboptimal and} {generally} less {effective} at instilling the knowledge of the expert to the agent \cite{hadfield2016cooperative}. {As argued by \cite{hadfield2016cooperative}, the value alignment problem  {with cooperative agents} is more appropriately viewed as an interactive decision-making process.} {The proposed formalism, termed CIRL,} is formulated as a two-player game of partial information with a common reward  function\footnote{{Also known as a \emph{team} problem \cite{marschak1972economic}.}}. Due to the special structure of CIRL, the problem can be transformed into a \emph{partially observable Markov decision process} (POMDP), see \cite{nayyar2013decentralized,hadfield2016cooperative}, allowing for single-agent RL algorithms to be applied. Further improvements in {computational} efficiency can be achieved by exploiting the fact that the expert expects the agent to {respond optimally} \cite{malik2018efficient}.

Inverse RL under a non-cooperative setting has not {received as much attention as its cooperative counterpart}. {A recent collection of work on multi-agent IRL (MA-IRL) }  \citep{lin2018multiagent,wang2018competitive,lin2019multi} addresses the problem of IRL in stochastic games {with multiple (usually more than two) agents}. {Distinct} from our N-CIRL setting, {MA-IRL} aims to recover the reward function of multiple agents under the assumption that the demonstrations are generated from the \emph{Nash equilibrium} strategies. Moreover, in the MA-IRL formalism, agents  {determine} their strategies based on the inferred reward function, \emph{i.e.}, regarding the inferred reward as some fixed ground truth. In contrast, under our N-CIRL setting, {  only one agent is unaware of the true reward function.} Furthermore, {the} goal of the less informed player  {in N-CIRL} goes beyond just \emph{inferring} the {true} reward function{; its ultimate goal is to determine an} optimal strategy against {a} worst-case opponent  {who possesses a private reward}. 

{From a game {theoretic} perspective, the N-CIRL formalism can be viewed as a stochastic dynamic game with asymmetric information, see \citep{cardaliaguet2009stochastic,nayyar2012dynamic,nayyar2014common,bacsar2014stochastic} and references therein.  In particular, N-CIRL {lies within the class of }games with one-sided incomplete information \citep{rosenberg1998duality,sorin2003stochastic,renault2006value,horner2010markov,horak2017heuristic}. This type of game allows for a simplified belief and allows the game to be decomposed into a sequence of single-stage games \cite{rosenberg1998duality,horak2017heuristic}. In particular, our N-CIRL formalism can be recognized as one of the {game} settings {discussed}  in \cite{rosenberg1998duality}, {in which a \emph{dual game} was formulated to solve for the less informed player's strategy.} {Our  algorithm for computing the defender's strategy is built upon this formulation, and can be viewed as one way to approximately solve the dual game. }

\section{Non-Cooperative Inverse Reinforcement Learning}\label{sec:model}
In this section, we {introduce} the N-CIRL formalism and describe its information structure. {As will be shown, the information structure of N-CIRL admits compact information states for each player.}

\subsection{N-CIRL Formulation}\label{subsec:Formulation}

The N-CIRL formalism is {modeled as} a two-player zero-sum Markov game with one-sided incomplete information. In particular, the attacker knows the true reward function, while the defender does not. In the context of the cyber security setting of this paper, the reward function is assumed to be parameterized by an intent parameter that is only known to the attacker. The N-CIRL formalism is described by the tuple
$\langle \S, \{\A,\D\}, \T(\cdot \mid \cdot, \cdot,\cdot), \{\Theta,R(\cdot,\cdot,\cdot,\cdot ;\cdot)\}, \mathcal{P}_0(\cdot, \cdot), \gamma \rangle$, where
\begin{itemize}
    \setlength\itemsep{-0.2em}
    \item $\S$ is the {finite} set of states; $s\in\S$.
    \item $\A$ is the {finite} set of actions for the attacker $A$; $\aA \in \A$.
    \item $\D$ is the {finite} set of actions for the defender $D$; $\aD \in \D$.
    \item {$\T(s' \mid s, a, d)$} is the conditional distribution of the next state $s'$ given current state $s$ and actions $a$, $d$. 
    \item $\Theta$ is the {finite} set of intent parameters that parameterize the reward function; the true intent parameter $\theta\in\Theta$ is only observed by the attacker.
    \item {$R(s,a,d,s';\theta)$ is the parameterized reward function that maps the current state $s \in\S$, actions $(a,d)\in\A\times\D$, next state $s' \in \S$}, and parameter $\theta \in \Theta$ to a reward for the attacker.  
    \item $\mathcal{P}_0$ is the distribution over the initial state $s_0$ and the true reward parameter $\theta$, assumed to be common knowledge between $A$ and $D$.
    \item $\gamma \in {[0,1)}$ is the discount factor.
\end{itemize}

The game proceeds as follows. Initially, a state-parameter pair $(s_0, \theta)$ is sampled from the prior distribution $\mathcal{P}_0$. The state $s_0$ is publicly observed by both players, whereas the intent parameter $\theta$ is only observed by the attacker.\footnote{The intent parameter $\theta$ is further assumed to be fixed throughout the problem.} For each stage, the attacker and the defender act simultaneously, choosing actions $a\in\A$ and $d\in\D$. Note that the action sets may be state-dependent, \emph{i.e.}, $a\in\A(s)$, $d\in\D(s)$. Given both actions, the current state {$s$} transitions to a successor state {$s'$} according to the transition model {$\T(s' \mid s, a, d)$}. The attacker receives a  {bounded} reward {$R(s, a, d, s';\theta)$}; the defender receives the reward {$-R(s, a, d, s';\theta)$} (incurs a cost {$R(s, a, d, s';\theta)$}). Neither player observes rewards during the game. {Before each subsequent stage, both players are informed of the successor state $s'$ and the actions from the previous stage}. While both players are aware of the {current} state, only the attacker is aware of the true intent parameter $\theta\in\Theta$. {This results in the defender possessing incomplete information, requiring it to maintain a belief over the true intent parameter.} {The goals of the attacker and the defender are to maximize and minimize the expected $\gamma$-discounted accumulated reward induced by $R$, respectively.}

\subsection{The Information Structure of N-CIRL}\label{subsec:info_structure}
 
{The N-CIRL formalism falls within the class of partially observable stochastic games \cite{hansen2004dynamic}}. In such games, perfect recall ensures that behavioral strategies, \emph{i.e.}, strategies that mix over actions, are outcome equivalent to mixed strategies, \emph{i.e.}, strategies that mix over pure strategies \cite{kuhn1953extensive}. As a result, players in N-CIRL can restrict attention to behavioral strategies, defined for each stage $t$ as $\pi^A_t: \H_t^A \to \Delta(\A)$ and $\pi^D_t: \H_t^D \to \Delta(\D)$, where $\H_t^A$ (resp. $\H_t^D$) represents the space of information available to the attacker (resp. defender) at stage $t$ and $\Delta(\A)$, $\Delta(\D)$ represent distributions over actions. Given any realized information sequences $I_t^A\in\H_t^A$ and $I_t^D\in\H_t^D$, represented as
\begin{align*}
I_t^A = \big(s_0,\theta,a_0,d_0,\ldots,a_{t-1},d_{t-1},s_t\big), \quad
I_t^D = \big(s_0,a_0,d_0,\ldots,a_{t-1},d_{t-1},s_t\big)
\end{align*}
the defender's information is always contained within the attacker's information for any stage $t$, \emph{i.e.}, there is one-sided incomplete information. Furthermore, note that the attacker has complete information (it knows everything that has happened in the game); its information at stage $t$ is the full history of the game at $t$, denoted by $I_t = \big(s_0,\theta,a_0,d_0,\ldots,a_{t-1},d_{t-1},s_{t}\big)\in\H_t=\H_t^A$.

{\bf Information States.} In general, games of incomplete information require players to reason over the entire {belief hierarchy}, that is, players' decisions not only depend on their beliefs on the state of nature, but also on the beliefs on others' beliefs on the state and nature, and so on \cite{mertens1985formulation,hansen2004dynamic}. 
Fortunately, players do not need to resort to this infinite regress in games of one-sided incomplete information. 
Instead, each player is able to maintain a compact state of knowledge, termed an \emph{information state}, that is sufficient for making optimal decisions. The more informed player maintains a pair consisting of the observable state and a distribution over the private state \cite{rosenberg1998duality,sorin2003stochastic}. 
The less informed player, through construction of a \emph{dual game} (discussed in Sec. \ref{ssec:def_strat}), maintains a pair consisting of the observable state and a vector (in Euclidean space) of size equal to the number of private states \cite{de1996repeated,rosenberg1998duality}. In the context of N-CIRL, the attacker's information state at each stage is a pair, denoted by $(s,b)$, in the space $\S\times\Delta(\Theta)$ whereas the defender's information state at each stage (of the dual game) is a pair, denoted by $(s,\zeta)$, in the space $\S\times\Rb^{|\Theta|}$.

\section{Solving N-CIRL} 
The theoretical results used to solve the CIRL problem \cite{nayyar2013decentralized,hadfield2016cooperative} do not extend to the N-CIRL setting. As outlined in \cite{hadfield2016cooperative}, the form of  CIRL allows one to convert the problem into a centralized control problem \cite{nayyar2013decentralized}.
The problem can then be solved using {existing} techniques from reinforcement learning. 

{In N-CIRL, such a conversion to a {centralized control problem} is not possible; one is instead faced with a dynamic game. As we show in this section, the one-sided incomplete information allows one to recursively define both the value of the game and the attacker's strategy. One can further recursively define the defender's strategy via the construction and sequential decomposition of a dual game. The two {recursive formulas} permit the development of a computational procedure, based on linear programming, for approximating both players' strategies.

\subsection{Sequential Decomposition}\label{ssec:valuebackup}

Solving a game involves finding strategies for all players, {termed} a \emph{strategy profile}, such that the resulting interaction is in (Nash) equilibrium. A strategy profile for N-CIRL is defined as follows.
\begin{defn}[Strategy Profile]
A strategy profile, denoted by $(\sigma^A,\sigma^D)$, is a pair of strategies $\sigma^A = (\pi_0^A,\pi_1^A,\ldots)$ and $\sigma^D = (\pi_0^D,\pi_t^D,\ldots)$, where $\pi_t^A$ and $\pi_t^D$ are behavioral strategies as defined in Sec. \ref{subsec:info_structure}.
\end{defn}
A simplification of behavioral strategies {are} strategies that only depend on the most recent information rather than the entire history. These strategies, termed \emph{one-stage strategies}, are defined below. 
\begin{defn}[One-Stage Strategies]
The one-stage strategies of the attacker and defender are denoted by $\bar\pi^A: \S\times\Theta\to\Delta(\A)$ and $\bar\pi^D: \S \to \Delta(\D)$, respectively. The pair $(\bar\pi^A,\bar\pi^D)$ is termed a one-stage strategy profile. 
\end{defn}

{Due to the information structure of N-CIRL, the {attacker's} information state $(s,b)$ can be updated using one-stage strategies instead of the full strategy profile. In fact, as illustrated by Lemma \ref{lem:beliefupdate}, the update of the attacker's information state only depends on the attacker's one-stage strategy $\bar\pi^A$, the attacker's action $a$, and the successor state $s'$.}

\begin{lemma}[Information State Update]\label{lem:beliefupdate}
{Given the attacker's one-stage strategy profile $\bar\pi^A$, the current attacker's information state $(s,b)\in\S\times\Delta(\Theta)$, the attacker's action $a\in\A$, }and the successor state $s'\in\S$, the attacker's updated information state is $(s',b')\in\S\times\Delta(\Theta)$ where the posterior $b'$ is computed via the function {$\tau:\S\times\Delta(\Theta)\times\A\to \Delta(\Theta)$}, defined elementwise as
\begin{align}\label{eq:beliefupdate}
	b'(\vartheta) = {\tau_{\vartheta}(s,b,a)}=
	\frac{\bar\pi^A(a \mid s, \vartheta)b(\vartheta)}{\sum\limits_{\vartheta' \in \Theta}\bar\pi^A(a \mid s, \vartheta')b(\vartheta')}.
\end{align}
\end{lemma}
 
{The {attacker's} information state $(s,b)$ also leads to the following definition of the value function   $v:\S\times\Delta(\Theta)\to\Rb$  of the game 
\begin{align*}
v(s,b)=\max_{\sigma^A}\min_{\sigma^D}~~\E\bigg[\sum_{t\geq 0}\gamma^t R(s_t,a_t,d_t,s_{t+1};\theta)\biggiven s_0=s,\theta\sim b(\cdot)\bigg],
\end{align*}
which denotes the minimax accumulated reward if the {initial state is} $s_0=s$, and the belief over $\Theta$ is $b$. 
Note that the value exists as {all spaces in} the game {are finite} and the discount factor {lies in $[0,1)$} \citep{rosenberg1998duality}.} The value function  $v$ can be computed recursively via a sequential decomposition. In fact, it is given by the fixed point of a value backup operator $[Gv](s,b)$, as illustrated in Proposition \ref{prop:seqdecomp_primal} below. %
{To differentiate from the {dual} game to be introduced in Sec. \ref{ssec:def_strat}, we refer to the original N-CIRL game as the \emph{primal} game and $[Gv](s,b)$ as the \emph{primal} backup operator.}

\begin{prop}[Sequential Decomposition of  Primal Game]\label{prop:seqdecomp_primal}
	The primal game can be sequentially decomposed into a sequence of single-stage games. {Specifically, the primal value function $v$ satisfies the following recursive formula
	\begin{align}\label{eq:valuebackup}
	v(s,b)=[Gv](s,b) = \max_{\bar\pi^A} \min_{\bar\pi^D} \big\{g_{\bar\pi^A, \bar\pi^D}{(s,b)} + \gamma V_{\bar\pi^A, \bar\pi^D}(v;{s,b}) \big\}
	\end{align}
	where   $[Gv](s,b)$ is referred to as  \emph{the primal value backup operator},} and $g_{\bar\pi^A, \bar\pi^D}{(s,b)}$, $V_{\bar\pi^A, \bar\pi^D}{(v;s,b)}$ correspond to the instantaneous reward and the expected value of the continuation game, respectively, defined as
	\begin{align}
	g_{\bar\pi^A, \bar\pi^D}{(s,b)} &= \sum_{a, d, s', \vartheta} b(\vartheta) \bar\pi^A(a \mid s, \vartheta) \bar\pi^D(d \mid s) \T(s' \mid s, a, d) R(s, a, d, s'; \vartheta)\label{eq:ins_reward}\\
	V_{\bar\pi^A, \bar\pi^D}(v{;s,b}) &= \sum_{a, d, s', \vartheta} b(\vartheta)\bar\pi^A(a \mid s, \vartheta) \bar\pi^D(d \mid s) \T(s' \mid s, a, d) v(s',b')\label{eq:con_reward}
	\end{align}
	where $b'$ represents the posterior distribution on $\Theta$ as computed by Eq. \eqref{eq:beliefupdate}.
\end{prop}

For purposes of constructing an algorithm, we need to establish some properties of the backup operator defined in Proposition \ref{prop:seqdecomp_primal}. The following lemma ensures that each application of the primal value backup yields a closer approximation of the value of the game. 

\begin{lemma}[Contraction of Primal Backup Operator]\label{lem:convergebackup_primal}
The {primal} value backup operator $[Gv](s,b)$, defined in Eq. \eqref{eq:valuebackup}, is a contraction mapping. {As a result, iterating the operator  converges to the value of the primal game that solves the fixed point equation \eqref{eq:valuebackup}.}
\end{lemma}  

{Though conceptually correct, iterating the backup operator $[Gv](s,b)$ exactly does not lead to a computationally tractable algorithm, as the belief $b$ lies in a continuous space with an infinite cardinality. Thus, {an} approximate value iteration algorithm is required for solving the fixed point equation \eqref{eq:valuebackup}. We will address this computational challenge in Sec. \ref{subsec:strate_comp}. 

Another challenge in solving N-CIRL is that the fixed point problem{, given by Eq.}  \eqref{eq:valuebackup}, cannot be solved by the defender. In fact, as pointed out in \cite[Sec. 1.2]{rosenberg1998duality}, if the defender is unaware of the attacker's strategy, it cannot form the posterior on $\Theta$. {The following section discusses the formulation of an auxiliary game to address this challenge.}}

\subsection{The Defender's Strategy} \label{ssec:def_strat}

{As shown by \cite{de1996repeated,rosenberg1998duality}, the defender's equilibrium strategy can be determined by construction of a dual game, characterized by a tuple $\langle \S, \{\A,\D\}, \T(\cdot \mid \cdot, \cdot,\cdot), \{\Theta,R(\cdot,\cdot,\cdot,\cdot ;\cdot)\}, \zeta_0, \mathcal{P}_0^{\S}(\cdot), \gamma\rangle$. Note that the sets $\S,\A,\D,\Theta$, the reward function $R(\cdot,\cdot,\cdot,\cdot ;\cdot)$, the discount factor $\gamma$, and the  state transition distribution $\T$ are identical to those in the primal game. {The quantity} $\zeta_0\in\Rb^{|\Theta|}$ is the parameter of the dual game, $\mathcal{P}_0^{\S}(\cdot)\in\Delta(\S)$ is the initial distribution of the state $s_0$, which is obtained by marginalizing   $\mathcal{P}_0(\cdot,\cdot)$ over $\theta$, \emph{i.e.}, $\mathcal{P}_0^{\S}(s)=\sum_{\theta\in\Theta}\mathcal{P}_0(s,\theta)$. The dual game proceeds as follows: at the initial stage, $s_0$ is sampled from $\mathcal{P}_0^{\S}$ and revealed to both players, the attacker \emph{chooses} some $\theta\in\Theta$; then the game is played identically as the primal one, namely, both players choose actions $a\in\mathcal{A}$ and $d\in\mathcal{D}$ simultaneously, and the state transitions from $s$ to $s'$ following $\T(\cdot\mid s, a, d)$.  {Both players are then informed of the chosen actions and the successor state $s'$.} {Furthermore, a} reward of $R(s,a,d,s';\theta)+\zeta_0(\theta)$ is received by the attacker (thus $-R(s,a,d,s';\theta)-\zeta_0(\theta)$ {is incurred} by the defender). Note that the value of $\theta$ is \emph{decided} and only known by the attacker, instead of being drawn from some probability distribution. This is one of the key differences from the primal game. 
 
The value function of the dual game, denoted by $w:\S\times\Rb^{|\Theta|}\to\Rb$, is defined as the maximin $\gamma$-discounted accumulated reward received by the attacker, if the state starts from some $s_0=s\in \S$ and the game parameter {is} $\zeta_0=\zeta\in \Rb^{|\Theta|}$. {The} value $w(s,\zeta)$ exists since the dual game is finite \citep{rosenberg1998duality}. Similarly as in Proposition \ref{prop:seqdecomp_primal}, the dual game value function $w$ also satisfies a recursive formula,  as formally stated below.}
 
\begin{prop}[Sequential Decomposition of  Dual Game]\label{prop:seqdecomp_dual}
	The dual game can be decomposed into a sequence of single-stage games. {Specifically, the dual value function $w$ satisfies the following recursive formula
	\begin{align}\label{eq:valuebackup_dual}
w(s,\zeta)=[Hw](s, \zeta) = & \min_{\bar\pi^D\!,\,\xi}\max_{\mu} \big\{\hD(s, \zeta) + \gamma\WD(w,\xi;s)\big\}
\end{align}
	where $[Hw](s, \zeta)$ is referred to as the \emph{dual value backup operator}, $\bar\pi^D(\cdot| s)\in\Delta(\Theta)$, $\xi \in \mathbb{R}^{\S \times \A \times \Theta}$ {are decision variables} with $\xi_{a,s}\in\Rb^{|\Theta|}$ the $(a,s)$th vector of $\xi$, $\mu \in \Delta(\A \times \Theta)$.} {Moreover}, $\hD(s, \zeta)$ and $\WD(w,\xi;s)$ 
	are defined as
	\begin{align}
\hD(s, \zeta) &:= \sum_{a, \vartheta}\mu(a, \vartheta) \Big(\zeta(\vartheta) + \sum_{d, s'}\bar\pi^D(d \mid s)\T(s'\mid s, a, d)R(s, a, d, s';\vartheta)\Big),\label{eq:ins_reward_dual}\\
\WD(w,\xi;s) &:= \sum_{a, d, s', \vartheta}\mu(a, \vartheta)\bar\pi^D(d \mid s)\T(s' \mid s, a, d)\big(w(s', \xi_{a, s'})-\xi_{a, s'}(\vartheta)\big).\label{eq:con_reward_dual} 
\end{align}
\end{prop}

{The recursive formula above  allows for a stage-wise calculation of the defender's strategy $\bar\pi^D$. In particular, by \cite{rosenberg1998duality}, the defender's equilibrium strategy obtained from the dual game is indeed its equilibrium strategy for the primal {game}. Moreover, the pair $(s,\zeta)$ is indeed the information state of the defender in the dual game. More importantly, the formula {of Eq.}\eqref{eq:valuebackup_dual} does not involve the update of the belief $b$, as opposed to {Eq.} \eqref{eq:valuebackup}. Instead, the vector $\xi$ plays the similar role as the updated belief $b'$, which can be \emph{calculated} by the defender as a decision variable. This way, as the defender plays the N-CIRL game, it can calculate its equilibrium  strategy by only observing the attacker's \emph{action} $a$ and the successive state $s'$, with no need to know the attacker's \emph{strategy}.  Besides, the defender's strategy $\bar\pi^D$, the  decision variable $\mu\in\Delta(\A\times\Theta)$ in {Eq. }\eqref{eq:valuebackup_dual}, is essentially the attacker's strategy in the dual game, \emph{i.e.}, given $\zeta$, the attacker has the flexibility to choose both $\theta\in\Theta$ and action $a\in\A$. 
	
{The} remaining issue is to solve the dual game by solving the fixed point equation \eqref{eq:valuebackup_dual}. As with the primal counterpart, the dual value backup operator is also {a contraction mapping}, as shown below. }

\begin{lemma}[Contraction of Dual Backup Operator]\label{lem:convergebackup_dual}
The value backup operator $[Hw](s,\zeta)$, defined in Eq. \eqref{eq:valuebackup_dual}, is a contraction mapping. {As a result, iterating the operator  converges to the dual game value that solves the fixed point equation \eqref{eq:valuebackup_dual}.}
\end{lemma}

{By Lemma \ref{lem:convergebackup_dual}, 
the defender's strategy can be obtained by 
iterating the backup operator. However, as $\zeta$ and $\xi$ both lie in continuous spaces, such an {iterative} approach is not computationally tractable. This motivates the approximate value iteration algorithm to be introduced next.}

\subsection{Computational Procedure}\label{subsec:strate_comp}

One standard algorithm for computing strategies in single-agent partially observable settings is the \emph{point-based value iteration} (PBVI) algorithm \cite{pineau2003point}. The standard PBVI algorithm approximates the value function, which is convex in the beliefs, using a set of $\alpha$-vectors. While the value functions $v$ and $w$ in N-CIRL also have desirable structure (as shown in \cite{rosenberg1998duality}, $v:\S\times\Delta(\Theta)\to\Rb$ is concave on $\Delta(\Theta)$ and $w:\S\times\Rb^{|\Theta|}\to\Rb$ is convex on $\Rb^{|\Theta|}$ for each $s\in\S$), the standard PBVI algorithm does not directly apply to N-CIRL. The challenge arises from the update step of the $\alpha$-vector set in PBVI, which requires carrying out   {one step of the} Bellman update \cite{pineau2003point}, for every action-observation pair of the agent. The corresponding Bellman update in N-CIRL is the primal backup operator in Eq. \eqref{eq:valuebackup}, which requires knowledge of the defender's strategy, something that is  unknown to the attacker.

To address {this challenge}, we develop a modified version of PBVI, termed \emph{non-cooperative PBVI} (NC-PBVI), in which the value functions $v$ and $w$ are approximated by a set of \emph{information state-value pairs}, \emph{i.e.,} $((s,b), v)$ and $((s,\zeta), w)$, instead of $\alpha$-vectors. Importantly, updating the sets only requires evaluations at individual information states, avoiding the need to know {the opponent's} strategy. 

The evaluations can be approximated using linear programming. Specifically, to approximate the value function of the primal game, NC-PBVI updates the value at {a given} attacker's information state $(s,b)$ by solving the primal backup operator of Eq. \eqref{eq:valuebackup}.  Using a standard reformulation, the minimax operation in the primal backup operator can be approximately solved via a linear program, denoted by $P_{A}(s, b)$. Similarly, one can approximately solve the dual game's fixed point equation, Eq. \eqref{eq:valuebackup_dual}, at a given defender's information state $(s, \zeta)$ via another linear program, denoted by $P_{D}(s, \zeta)$. For notational convenience, define $T_{sad}(s') = \T(s'\mid s,a,d)$, $P_{sad}^\vartheta = \sum_{s'}\T(s'\mid s,a,d)R(s,a,d,s';\vartheta)$, $A_{s\vartheta}(a) = \bar\pi^A(a\mid s,\vartheta)b(\vartheta)$, and $D_s(d) = \bar\pi^D(d\mid s)$. 
The two linear programs, $P_{A}(s, b)$ and $P_{D}(s, \zeta)$, are given below. 

\begin{figure}[h!]
	\begin{minipage}[t]{.5\textwidth}
		\begin{align*}
		&\hspace{-2.3em}\max\limits_{\substack{[A_{s\vartheta}(a)] \geq 0, [V_{ds'}], \\ [b_{ds'}(\vartheta)]\geq 0, V}} \quad V \hspace{5em} \mybox{$P_{A}(s, b)$}\\
		\text{s.t. } &V \leq \sum\limits_{a, \vartheta} A_{s\vartheta}(a) P_{sad}^\vartheta + \gamma\sum\limits_{s'}V_{ds'}\quad\forall \,d \\
		& b_{ds'}(\vartheta) = \sum\limits_{a}A_{s\vartheta}(a) T_{sad}(s')\quad\forall \,d, s', \vartheta \\
		& V_{ds'} \leq \Upsilon_v\big(\Y^A_s, \W^A_s, b_{ds'}\big)\quad\forall \,d, s'\\
		& \sum\limits_{a} A_{s\vartheta}(a) = b(\vartheta)\quad\forall \,\vartheta 
		\end{align*}
	\end{minipage}\hspace{0.25em}
	\vline\hspace{0.25cm}
	\begin{minipage}[t]{.5\textwidth}
		\begin{align*}
		&\hspace{-3.5em}\hspace{1em}\min_{\substack{[D_s(d)] \ge 0, [W_{as'}], \\ [\lambda_{as'}(\vartheta)]\ge 0, W}} \quad W \hspace{5em} \mybox{$P_{D}(s, \zeta)$}\\
		\text{s.t. } & W \geq \zeta(\vartheta) + \sum_dD_s(d)P_{sad}^\vartheta\\
		&\hspace{1.25em} + \gamma\sum_{s'}\big(W_{as'}- \lambda_{as'}(\vartheta)\big)\quad\forall \,a, \vartheta\\
		&W_{as'} \geq \Upsilon_w\big(\Y^D_s, \W^D_s,\lambda_{as'}\big)\quad\forall \,a, s' \\
		\nonumber&\sum_{d}D_s(d) = 1
		\end{align*}
	\end{minipage}
\end{figure}
 
{The objective functions of the two linear programs estimate the values of the primal and dual games. The decision variables $[A_{s\vartheta}],[V_{ds'}],[b_{ds'}(\vartheta)]$ in $P_{A}(s, b)$ are used to find the attacker's strategy, the continuation value of the (primal) game, and the updated belief, respectively. Similarly, $[D_s(d)],[W_{as'}],[\lambda_{as'}(\vartheta)]$ in $P_{D}(s, \zeta)$ are used to find the defender's strategy, the continuation value of the dual game, and the updated parameter $\zeta$ for the dual game, respectively.}   
The first constraint in $P_{A}(s, b)$ encodes the defender's best response, replacing the minimization over $\bar\pi^D$ in Eq. \eqref{eq:valuebackup}. Similarly, the first constraint in $P_{D}(s, \zeta)$ replaces the maximization over $\mu$ in Eq. \eqref{eq:valuebackup_dual}. The second and last constraints in $P_{A}(s, b)$ and the last constraint in $P_{D}(s, \zeta)$ enforce basic rules of probability. {The third constraint in $P_{A}(s, b)$ and the second constraint in $P_{D}(s, \zeta)$ provide the  {information state-value} approximations of the continuation value estimates $V_{ds'}$ and $W_{as'}$, respectively.} 

{Due to concavity (convexity) of the value function $v$ (resp. $w$), we use the \emph{sawtooth} function }\cite{hauskrecht2000value} as an information state-value approximation. In particular, a lower bound on the primal game's value function $v(s,b)$ is given by $\Upsilon_v\big(\Y^A_s, \W^A_s, b_{ds'}\big)$ whereas an upper bound of the dual game's value function $w(s,\zeta)$ is given by $\Upsilon_w\big(\Y^D_s, \W^D_s,\lambda_{as'}\big)$.\footnote{Note that $b_{ds'}$ is the vector consisting of $b_{ds'}(\vartheta)$ over all $\vartheta\in\Theta$ (similarly for $\lambda_{as'}$).} The set $\Y^A_s$ contains the belief-value pairs associated  with the beliefs that are non-corner points of the simplex over $\Theta$ for given $s$, whereas the set $\W^A_s$ contains the belief-value pairs associated  with the corner points of the simplex. Analogously, $\Y^D_s, \W^D_s$ represent  subsets of $\Rb^{|\Theta|}$ that contain the vectors with only  one, and  more than one, non-zero element, respectively. {Details of both $\Upsilon_v$ and $\Upsilon_w$ using sawtooth functions can be found in the pseudocode in Sec. \ref{sec:pbvi} in the Appendix. }

Lemma \ref{lem:sawtooth} ensures that the sawtooth constraints are linear in the decision variables of the respective problem{, which verifies the computational tractability of $P_{A}(s, b)$ and $P_{D}(s, \zeta)$. The proof of Lemma \ref{lem:sawtooth} can be found in Sec. \ref{sec:proof_append} in the appendix.}

\begin{lemma}\label{lem:sawtooth}
	By definitions of \emph{SAWTOOTH-A} and \emph{SAWTOOTH-D} in Algorithm \ref{alg:PBVI_NCIRL}, given $\Y^A_s, \W^A_s,\Y^D_s, \W^D_s$, constraints $V_{ds'} \leq \Upsilon_v\big(\Y^A_s, \W^A_s, b_{ds'}\big)$ and $W_{as'} \geq \Upsilon_w\big(\Y^D_s, \W^D_s,\lambda_{as'}\big)$ are both \emph{linear} in the decision variables $V_{ds'},b_{ds'}$ and $W_{as'},\lambda_{as'}$, respectively. 
\end{lemma}

Next, we numerically analyze the proposed algorithm in a cyber security environment.

\section{Experiment: Intrusion Response}\label{sec:exp}
A recent trend in the security literature concerns the development of automated defense systems, termed state-based intrusion response systems, that automatically prescribe defense actions in response to intrusion alerts \cite{miehling2015optimal,iannucci2016probabilistic,miehling2018pomdp}. Core to these systems is the construction of a model that describes the possible ways an attacker can infiltrate the system, termed a \emph{threat model}. Deriving a correct threat model is a challenging task and has a significant impact on the effectiveness of the intrusion response system. N-CIRL addresses one of the main challenges in this domain: the defender's uncertainty of the attacker's true intent. 

The threat model in our experiment is  {based on} an \emph{attack graph}, a common graphical model in the security literature \cite{ammann2002scalable}. {An} attack graph is represented by a directed acyclic graph $\G=(\N,\Es)$ where each node $n\in\N$ represents a system condition and each edge $e_{ij}\in\N\times\N$ represents an exploit. Each exploit $e_{ij}$ relates a \emph{precondition} $i$, the condition needed for the attack to be attempted, to a \emph{postcondition} $j$, the condition satisfied if the attack succeeds. Each exploit $e_{ij}$ is associated with a probability of success, $\beta_{ij}$, describing the likelihood of the exploit succeeding (if attempted). The state space $\S$ is the set of currently satisfied conditions (enabled nodes). For a given state $s\in\S$, the attacker chooses among exploits that have enabled preconditions and {at least one not yet enabled postcondition}. The defender simultaneously chooses which exploits to block for the current stage; blocked edges have a probability of success of zero for the stage in which they are blocked. The attacker's reward is $R(s,a,d,s';\theta) = r_e(s,s';\theta) - c_A(a) + c_D(d)$, where $s'$ is the updated state, $r_e(s,s';\theta)$ is the attacker's reward for any newly enabled conditions, and $c_A(a)$ and $c_D(d)$ are costs for attack and defense actions, respectively. The experiments are run on random instances of attack graphs; see some instances in Figure \ref{plot:value}. See Sec. \ref{sec:pbvi} for more details of the experimental setup.

As seen in Figure \ref{plot:value}, the strategies obtained from N-CIRL yield a lower attacker reward than the strategies obtained from MA-IRL. Empirically, this implies that the defender benefits more from interleaving learning and execution than the attacker. Even though the interleaved setting may provide more ground for the attacker to exercise deceptive tactics, \cite{rosenberg2000maxmin} states that in games of incomplete information, the more informed player ``cannot exploit its private information without revealing it, at least to some extent.'' In the context of our example, we believe that the performance gain of N-CIRL arises from the fact that the attacker can only deceive for so long; eventually it must fulfill its true objective and, in turn, reveal its intent. 

\begin{figure}[h!]
	\vspace{-0.5em}
	\includegraphics[width=\textwidth]{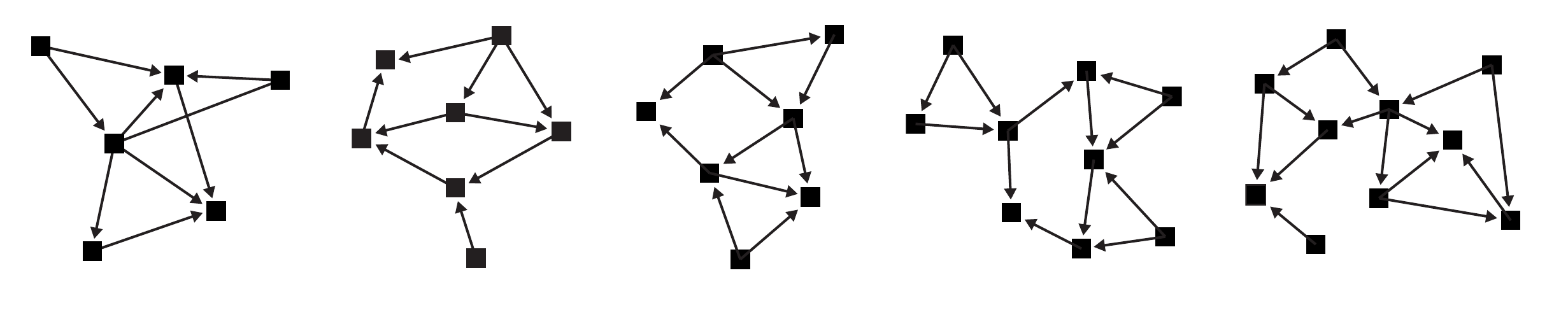}
	\includegraphics[height=4.25cm]{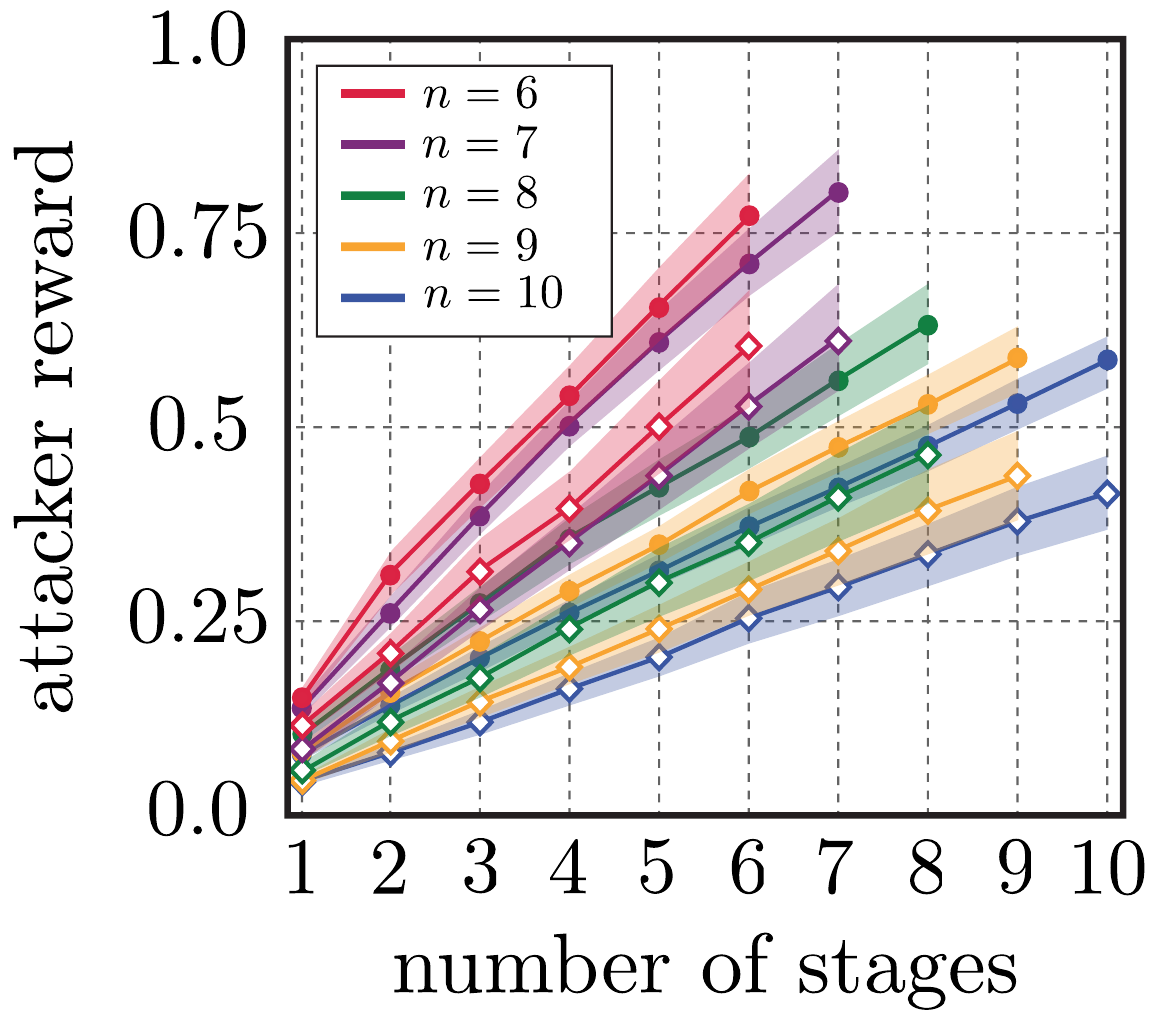}
	\includegraphics[height=4.25cm]{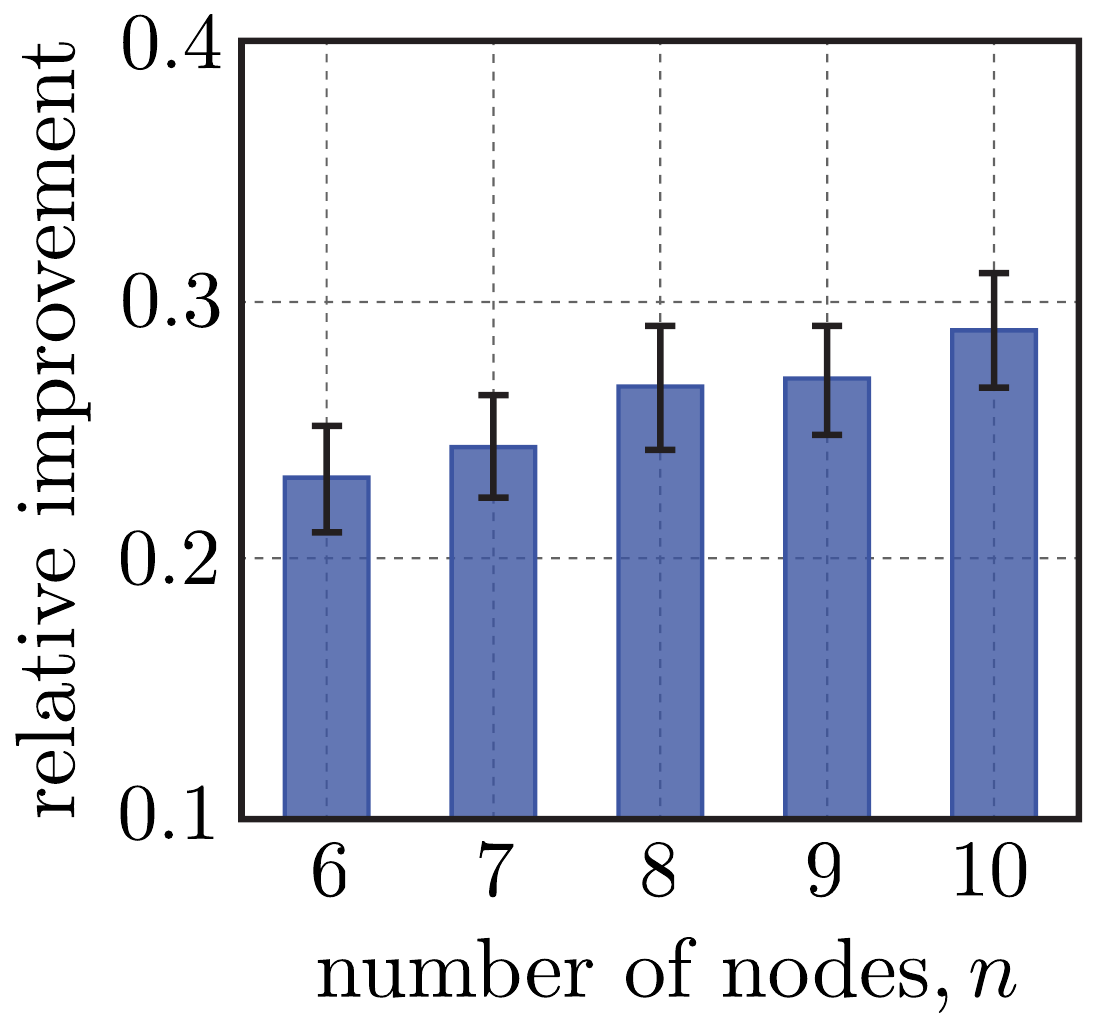}
	\includegraphics[height=4.25cm]{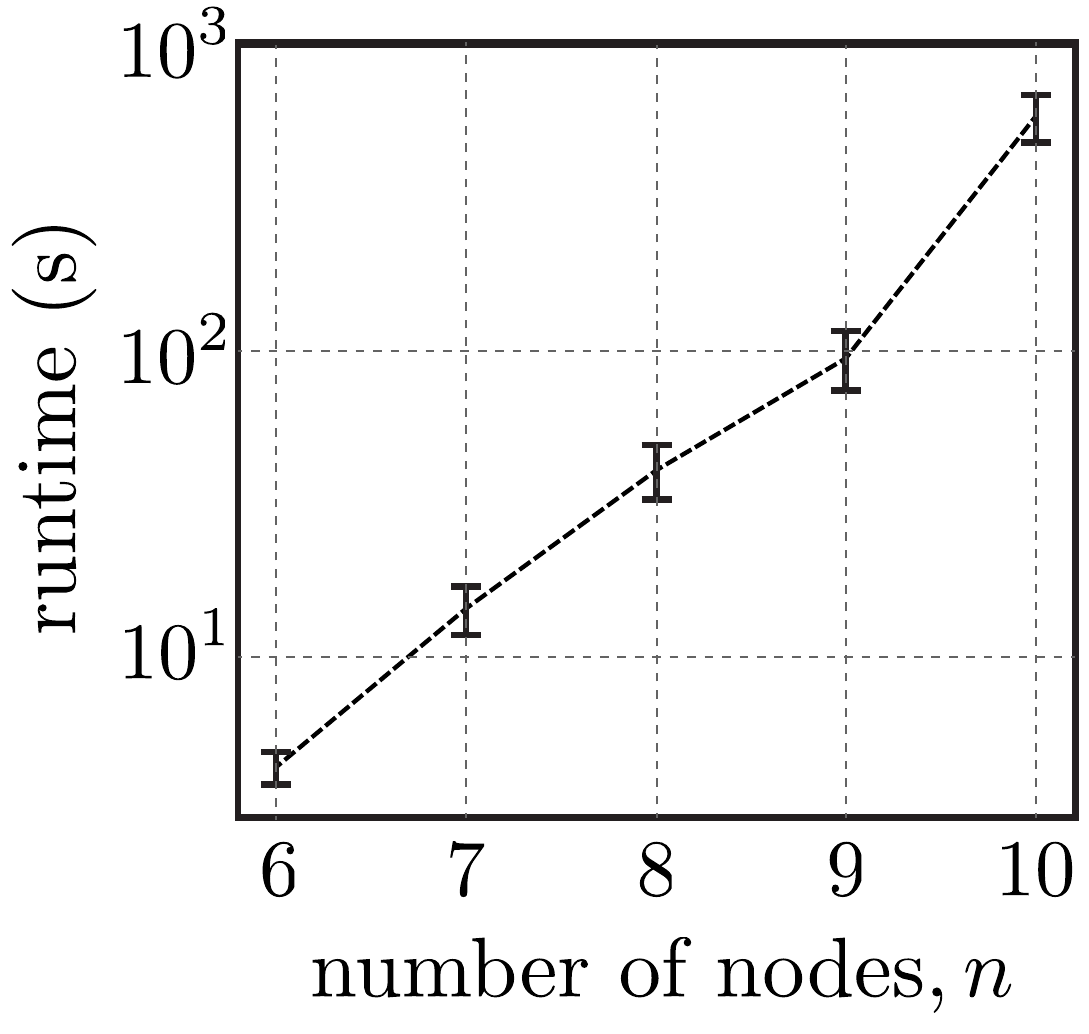}
	\caption{\emph{Top}: Instances of randomly generated graphs of sizes $n=6$ to $n=10$, with state cardinalities $|\S| = \{8, 24, 32, 48, 64\}$ and action cardinalities $|\A| = |\D| = \{54, 88, 256, 368, 676\}$. \emph{Bottom-left}: Attacker's average reward for each graph size $n$; filled/outlined markers represent the attacker's accumulated reward versus defense strategies obtained from MA-IRL/N-CIRL. \emph{Bottom-middle}: The average relative reduction of attacker reward in N-CIRL compared to MA-IRL as a function of $n$. \emph{Bottom-right}: Average runtime of NC-PBVI (in seconds) as a function of $n$.}
	\label{plot:value}
	\vspace{-1em}
\end{figure} 
 
\section{Concluding Remarks}
The goal of our paper was to introduce the N-CIRL formalism and provide some theoretical results for the design of learning algorithms in the presence of strategic opponents. The primary motivation for this work was cyber security, specifically, problems where the defender is actively trying to defend a network when it is uncertain of the attacker's true intent. Learning from past attack traces (\emph{i.e.}, equilibrium behavior/demonstrations) can lead to poor defense strategies, demonstrating that such approaches (\emph{e.g.}, MA-IRL) are not directly applicable for settings where rewards may change between demonstrations. Empirical studies illustrate that the defender can benefit by interleaving the learning and execution phases compared to just learning from equilibrium behavior (this is an analogous conclusion to the one found in CIRL that an interactive scenario can lead to better performance). The reason for this is that defense strategies computed using N-CIRL learn the intent adaptively through interaction with the attacker.

As shown in \cite{hadfield2016cooperative}, the value alignment problem is more appropriately addressed in a dynamic and cooperative setting. The cooperative reformulation converts the IRL problem into a decentralized stochastic control problem. In our paper, we have shown that the non-cooperative analog of CIRL, \emph{i.e.}, when agents possess goals that are misaligned, becomes a zero-sum Markov game with one-sided incomplete information. Such games are conceptually challenging due to the ability of agents to influence others' beliefs of their private information through their actions, termed \emph{signaling} in game theoretic parlance. We hope that the N-CIRL setting can provide a foundation for an algorithmic perspective of these games and deeper investigation into signaling effects in general stochastic games of asymmetric information.

\section*{Acknowledgements}
This work was supported in part by the US Army Research Laboratory (ARL) Cooperative Agreement W911NF-17-2-0196, and in part by the Office of Naval Research (ONR) MURI Grant N00014-16-1-2710.

\bibliographystyle{abbrv} 
\bibliography{ref}

\newpage

\appendix

~\\
\centerline{{\fontsize{16}{16}\selectfont \textbf{Supplementary Materials for ``Non-Cooperative Inverse}}}

\vspace{6pt} 
 \centerline{\fontsize{16}{16}\selectfont \textbf{
 Reinforcement Learning''}}

\section{Illustrative  Example}\label{ssec:example}
  
Consider a simple zero-sum patrolling game where a thief $A$ aims to steal valuables from a museum $m$ and a gallery $g$ that a security guard $D$ is {watching over}. The state space is defined as the set of possible locations of the players $(A,D)$, defined as $\S = \{(m, m), (m, g), (g, m), (g, g)\}$.   
The initial state is uniformly sampled from $\S$. The game is played over multiple stages, where at each stage of the game, $A$ and $D$ observe the current state $s\in\S$ and (simultaneously) choose to either stay at their respective locations or switch to the other location, \emph{i.e.}, $\A = \D =\{\text{stay, switch}\}$. The state deterministically transitions from $s$ to $s'$ based on $(a,d)$. If the state transitions to either $s'= (m, g)$ or $s' = (g, m)$, $A$ will successfully steal an item with probability $1$ and gain a reward of either $\theta$ if $s'= (m, g)$, or $1-\theta$ if $s' = (g, m)$, where $\theta\in[0,1]$ reflects $A$'s private preference that is unknown to $D$. If the state transitions to either $s' = (m, m)$ or $s' = (g, g)$, the presence of $D$ will lower $A$'s probability of success to $1/2$. Thus, $A$'s (expected) reward at each stage is 

\begin{align}\label{equ:def_reward}
R(s;\theta)  = \left\{\begin{array}{ll}   
\theta/2 &  \text{ if } s= (m, m)\\
\theta &  \text{ if } s= (m, g)\\
1-\theta &  \text{ if } s= (g, m)\\
(1-\theta)/2 &  \text{ if } s= (g, g)\\
\end{array}\right.
\end{align}

Under this setting, we compare the strategies obtained under the MA-IRL formalism, \emph{e.g.},  \cite{lin2018multiagent,wang2018competitive,lin2019multi}, to the strategies where $D$ can learn the intent adaptively under the N-CIRL formalism. In MA-IRL, $D$ has access to an attack log, which reflects that $A$ prefers the gallery twice as much as the museum, \emph{i.e.}, $\theta=1/3$. In an initial \emph{learning phase}, $D$ learns from previous equilibrium behavior\footnote{$D$ under MA-IRL is (generously) assumed to be able to \emph{perfectly recover} the parameter $\theta = 1/3$ during the learning phase. This is often not the case as $D$'s  inference using historical data can not be exactly accurate in practice. We make this assumption to favor MA-IRL as much as possible.}, and in the subsequent \emph{execution phase}, $A$ and $D$ play the game described above. In contrast, $D$ under N-CIRL has no prior knowledge of $A$'s intent and learns it from scratch through information that is revealed as the game unfolds. For purposes of this example, $A$'s preferences are assumed to flip in the execution phase, \emph{i.e.}, $\theta=2/3$. We now compare two strategies in a two-stage instance of the above game.

{\bf MA-IRL Strategies.}
For $\theta=1/3$, $D$ has a unique pure Nash equilibrium strategy that generates its next location to be $g$, independent of the initial state. However, since $A$'s true intent in the execution phase when $D$ actually participates is $\theta=2/3$, $A$'s pure Nash equilibrium strategies is to go to $m$. Such a combination of $A$ and $D$'s equilibrium strategies yields a next state of $(m,g)$, which results in a reward of $2/3$ for $A$. Hence, over the two-stage game (execution phase), the total reward for $A$ against $D$ under MA-IRL is $2\cdot 2/3 = 4/3$. 

{\bf N-CIRL Strategies.} 
$D$ has a uniform patrol strategy in the first stage (toss a fair coin to decide where to patrol). Since $A$  has intent parameter $\theta=2/3$ in the execution phase,  it will go to the museum according to $R(s;\theta=2/3)$, with state $(m,m)$ arising from the Nash equilibrium strategies. Hence, $A$'s expected reward in the first stage is $1/2\cdot1/2\cdot2/3+1/2\cdot2/3=1/2$. At the second stage, $D$ receives the observation that $A$ went to $m$, thus infers $\theta>1/2$.  With the new estimation of $A$'s intent, $D$ prefers to defend $m$, which results in a cost of  $-1/2\cdot\theta$ that is smaller than the cost of defending  $g$, which is $-\theta$. Therefore, the expected reward of $A$ at the second stage becomes $1/2\cdot2/3=1/3$, and the total reward in the two-stage game is $1/2+1/3=5/6<4/3$.  
 
The above example illustrates that there exist settings where the defender can incur {a lower} cost {by interleaving} learning and {execution}. 

\clearpage

\section{Proofs of Main Results}\label{sec:proof_append}

\subsection{Proof of Lemma \ref{lem:beliefupdate}}

\begin{proof}

Given a current state $s$ and distribution $b$ on $\Theta$, the posterior distribution is computed by conditioning on the new information (consisting of the actions $(a,d)$ and updated state $s'$) as 
\begin{align}
P(\theta=\vartheta \mid { s,  b,  a, d, s'}) &= \frac{P({s, b, a, d, s'}\mid \vartheta)P(\theta=\vartheta)}{\sum_{\vartheta'} P({s, b, a, d, s'}\mid \vartheta')P(\theta=\vartheta')} \label{equ:proof_lemma1P_1}\\
&= \frac{P(s\mid \vartheta)P(a,d\mid s,\vartheta)P({b,s'}\mid s,a,d,\vartheta)P(\theta=\vartheta)}{\sum_{\vartheta'} P({s, b, a, d, s'}\mid \vartheta')P(\theta=\vartheta')} \label{equ:proof_lemma1P_2}\\
&= \frac{P(s\mid \vartheta)P(a\mid s,\vartheta)P(d\mid s)P({s'}\mid s,a,d)P(b\mid s,\vartheta)P(\theta=\vartheta)}{\sum_{\vartheta'} P({s, b, a, d, s'}\mid \vartheta')P(\theta=\vartheta')} \label{equ:proof_lemma1P_3}\\
&= \frac{P(s\mid \vartheta)P(a\mid s,\vartheta)P(d\mid s)P({s'}\mid s,a,d)P(b\mid s,\vartheta) b(\vartheta) }{\sum_{\vartheta'}P(s\mid \vartheta')P(a\mid s,\vartheta')P(d\mid s)P({s'}\mid s,a,d)P(b\mid s,\vartheta') b(\vartheta')} \label{equ:proof_lemma1P_4}\\
&= \frac{P(s'\mid s, a, d)  P(d \mid s) P(a \mid s, \vartheta) b(\vartheta)}{P(s'\mid s, a, d) P(d \mid s) \sum_{\vartheta'} P(a \mid s, \vartheta') b(\vartheta')} \label{equ:proof_lemma1P_5}\\
&= \frac{P(a \mid s, \vartheta) b(\vartheta)}{\sum_{\vartheta'}P(a \mid s, \vartheta') b(\vartheta')},\label{equ:proof_lemma1P_6}
\end{align}
{where Eq. \eqref{equ:proof_lemma1P_1} follows Bayes rule, Eq. \eqref{equ:proof_lemma1P_2} uses definition of conditional probability, Eq. \eqref{equ:proof_lemma1P_3} uses the conditional independence of $a$ and $d$ given $s,\vartheta$, and that of $b$ and $s'$ given $s,a,d,\vartheta$, Eq. \eqref{equ:proof_lemma1P_4}  expands the denominator as in Eq. \eqref{equ:proof_lemma1P_3} and uses the fact that $P(\theta=\vartheta')=b(\vartheta')$,  Eq. \eqref{equ:proof_lemma1P_5} uses that $P(s\mid \vartheta)=P(b\mid s,\vartheta)=1$, and Eq. \eqref{equ:proof_lemma1P_6} follows by cancelling out $P(s'\mid s,a,d)$. 
 Note that the probabilities $P(s'\mid s,a,d),P(d\mid s),P(a\mid s,\vartheta)$ are all nonzero, as the calculation in Eqs.  \eqref{equ:proof_lemma1P_1}-\eqref{equ:proof_lemma1P_6} is only for those tuples of $(a,d,s')$ that have been realized.}
Recognizing that $P(a \mid s, \vartheta)=\bar\pi^A(a \mid s, \vartheta)$  yields the result. 
\end{proof}

\subsection{Proofs of Propositions \ref{prop:seqdecomp_primal} and \ref{prop:seqdecomp_dual}}

\begin{proof} The proofs of Propositions \ref{prop:seqdecomp_primal} and \ref{prop:seqdecomp_dual} are similar and are built upon the results of \cite{rosenberg1998duality}. In the language of \cite{rosenberg1998duality}, player 1 is the more informed player and player 2 is the less informed player. Let $I$ and $J$ denote the action sets of player 1 and 2, respectively, $K$ denotes the set of the states of the world, $X$ denotes the set of the  stochastic states, and $\lambda\in(0,1)$ is  the discount factor. Let $s\in\Delta(I)^K$ denote the strategy\footnote{The notation $\Delta(I)^K$ means all functions from $K$ to $\Delta(I)$.} of player 1 and $t\in\Delta(J)$ denote the strategy of player 2. To prove Proposition \ref{prop:seqdecomp_primal}, consider the following sequential decomposition from \cite[Proposition 6]{rosenberg1998duality}, 
\begin{align}
\nonumber v_{\lambda}(p,x) &= \max_{s\in\Delta(I)^K}\min_{t\in\Delta(J)}\bigg[\lambda\sum_{k\in K}\sum_{(i,j)\in I\times J}p^ks^k(i)A_{ij}^{kx}t(j)\\
&+(1-\lambda)\sum_{k\in K}\sum_{(i,j)\in I\times J}\sum_{y\in X}p^ks^k(i)t(j)q(x,i,j,y)v_{\lambda}(p_i,y)\bigg],\label{eq:primalRosenberg}
\end{align}
where superscripts denote indexing and $p_i$ denotes the Bayesian update defined elementwise by $p_i^k = \frac{p^ks^k(i)}{\sum_{l\in K}p^ls^l(i)}$. There are two modifications that need to be made to the recursive formula. First, compared to the reward function $A$ in \cite{rosenberg1998duality}, the reward function in N-CIRL additionally depends on the successor state $s'$. This dependence requires that the first term in the summation of Eq. \eqref{eq:primalRosenberg} also needs to take an expectation over $s'$. Second, the form of the payoff differs in N-CIRL; dividing Eq. \eqref{eq:primalRosenberg} by $\lambda$  yields an expression for $v_{\lambda}/\lambda$ which corresponds to value of the primal game $v$ in Proposition  \ref{prop:seqdecomp_primal}. Denoting $s,t,p,A,q,1-\lambda$ therein by $\bar\pi^A, \bar\pi^D,b,R,\T,\gamma$ in our notation system, respectively, yields the sequential decomposition of \eqref{eq:valuebackup} in Proposition \ref{prop:seqdecomp_primal}. 

To prove Proposition \ref{prop:seqdecomp_dual}, consider the sequential decomposition from \cite[Proposition 7]{rosenberg1998duality}, 
\begin{align}
\nonumber w_\lambda(\alpha,x) &= \inf_{t\in\Delta(J)}\inf_{\beta\in\mathbb{R}^{K\times I\times X}}\sup_{\pi\in\Delta(I\times K)}\sum_{i\in I}\sum_{k\in K}\pi(i,k)\bigg(\lambda t(j)A_{ij}^{kx}+\alpha^k\bigg)\\
\nonumber & -\sum_{k\in K}\sum_{(i,j)\in I\times J}\sum_{y\in X}(1-\lambda)\pi(i,k)\beta^k(i,y)t(j)q(x,i,j,y)\\
& +\sum_{i\in I}\sum_{y\in X}(1-\lambda)\bigg(\sum_{j\in J}t(j)q(x,i,j,y)\bigg)\pi(i)w_\lambda(\beta(i,y),y)\label{eq:dualRosenberg}
\end{align}
where $\pi(i) = \sum_{k\in K}\pi(i,k)$ and $\alpha\in\Rb^K$. First, note that the dual game in our case is finite ($I,J,K,X$ are finite in our problem) and hence has a value. As a result, the $\inf$ and $\sup$ in \cite[Propositions 7]{rosenberg1998duality} can be replaced by $\min$ and $\max$. Next, as in the proof of Proposition \ref{prop:seqdecomp_primal}, divide Eq. \eqref{eq:dualRosenberg} by $\lambda$ and denote $t,\alpha/\lambda,\beta/\lambda,\pi,\omega_{\lambda}/\lambda,1-\lambda$ therein by $\bar\pi^D,\zeta,\xi,\mu,w,\gamma$, respectively. Grouping terms, we arrive at the sequential decomposition of \eqref{eq:valuebackup_dual} in Proposition \ref{prop:seqdecomp_dual}.\end{proof} 

\subsection{Proof of Lemma \ref{lem:convergebackup_primal}}
\begin{proof}
	We prove this lemma by showing that the value backup operator $G$ for a given one-stage strategy profile $(\bar\pi^A, \bar\pi^D)$ is a contraction mapping with $\gamma \in [0,1)$. Let $\eta := (s, b)$ and $v, v'$ be value functions. By Blackwell's sufficiency theorem \cite{blackwell1965discounted}, $[Gv](\eta)$ is a contraction mapping if it satisfies i) monotonicity: $[Gv](\eta)\geq[Gv'](\eta), \text{ if }v(\eta)\geq v'(\eta)$ for any $\eta$, and ii) discounting: $[Gv](\eta) = [Gv'](\eta) + \gamma\varepsilon, ~\forall v(\eta) = v'(\eta) + \varepsilon$. To show monotonicity, assume $v(\eta) \geq v'(\eta)$ for all $\eta$; then we have
	\begin{align*}
		\sum\limits_{a, d, s', \vartheta} b(\vartheta)\bar\pi^A(a \mid s, \vartheta) \bar\pi^D(d \mid s) \T(s' \mid s, a, d)(v(\eta') - v'(\eta')) \geq 0 \quad \forall \eta
	\end{align*} 
	where $\eta'$ represents the updated attacker information state. Note that the instantaneous reward does not depend on $v$, $v'$, thus 
	\begin{align*} 
	[Gv](\eta)-[Gv'](\eta) &= \gamma \max\limits_{\bar\pi^A} \min_{\bar\pi^D} \big\{ V_{\bar\pi^A, \bar\pi^D}(v; \eta) - V_{\bar\pi^A, \bar\pi^D}(v'; \eta)\big\} \\
	&\hspace{-2em}= \gamma \max\limits_{\bar\pi^A} \min_{\bar\pi^D} \Big\{ \sum\limits_{a, d, s', \vartheta} b(\vartheta)\bar\pi^A(a \mid s, \vartheta) \bar\pi^D(d \mid s) \T(s' \mid s, a, d)(v(\eta') - v'(\eta'))\Big\} \\
	&\hspace{-2em} \geq 0.
	\end{align*}
	To show discounting, let $v(\eta) = v'(\eta) + \varepsilon$. Then
	\begin{align*}
	[Gv](\eta) &= \max\limits_{\bar\pi^A} \min_{\bar\pi^D} \bigg\{ \sum\limits_{a, d, s', \vartheta} b(\vartheta) \bar\pi^A(a \mid s, \vartheta) \bar\pi^D(d \mid s) \T(s' \mid s, a, d) R(s, a, d, s';\vartheta) \\
	&\hspace{8em} + \gamma \sum_{a, d, s', \vartheta}b(\vartheta)\bar\pi^A(a\mid s, \vartheta)\bar\pi^D(d \mid s) \T(s' \mid s, a, d) (v'(\eta') + \varepsilon) \bigg\}\\
	&= \max\limits_{\bar\pi^A} \min_{\bar\pi^D} \bigg\{ \sum\limits_{a, d, s', \vartheta} b(\vartheta) \bar\pi^A(a \mid s, \vartheta) \bar\pi^D(d \mid s) \T(s' \mid s, a, d) R(s, a, d, s';\vartheta) \\
	&\hspace{8em} + \gamma \sum_{a, d, s', \vartheta}b(\vartheta)\bar\pi^A(a\mid s, \vartheta)\bar\pi^D(d \mid s) \T(s' \mid s, a, d) (v'(\eta')) \bigg\}  + \gamma\varepsilon\\
	&=  [Gv'](\eta) + \gamma\varepsilon.
	\end{align*}
	
	Therefore, the one-stage value backup operator is a contraction mapping for a given one-stage strategy profile $(\bar\pi^A, \bar\pi^D)$. 
\end{proof}

\subsection{Proof of Lemma \ref{lem:convergebackup_dual}}}
\begin{proof}
	Let $w, w'$ be value functions. As in the proof of Lemma \ref{lem:convergebackup_primal}, we show the monotonicity and discounting properties, and then invoke Blackwell's sufficiency theorem. Assume $w(\xi, s) \geq w'(\xi, s)$ for all $s, \xi$; then
	\begin{align*}
	\sum\limits_{a, d, s', \vartheta} \mu(a, \vartheta) \bar\pi^D(d \mid s) \T(s' \mid s, a, d)(w(\xi_{a, s'}, s') - w'(\xi_{a, s'}, s')) \geq 0 \quad \forall s, \xi
	\end{align*} 
	Note that the instantaneous reward does not depend on $w$, $w'$, and thus 
	\begin{align*} 
	&\hspace{1.5em}[Hw](s, \zeta)-[Hw'](s, \zeta) \\
	&= \gamma \min\limits_{\bar\pi^D, \xi} \max_{\mu} \big\{ \WD(w, \xi; s) - \WD(w', \xi; s)\big\} \\
	&= \gamma \min\limits_{\bar\pi^D, \xi} \max_{\mu} \big\{\sum\limits_{a, d, s', \vartheta} \mu(a, \vartheta) \bar\pi^D(d \mid s) \T(s' \mid s, a, d)(w(\xi_{a, s'}, s') - w'(\xi_{a, s'}, s'))\big\} \\
	&\geq 0.
	\end{align*}
	
	To show discounting, let $w(\xi, s) = w'(\xi, s) + \varepsilon$ for all $s, \xi$. Then
	\begin{align*}
	[Hw](s, \zeta) &= \min_{\bar\pi^D, \xi}\max_{\mu} \bigg\{ \sum_{a, \vartheta}\mu(a,\vartheta) \bigg(\zeta(\vartheta) + \sum_{d, s'}\bar\pi^D(d \mid s)\T(s'\mid s, a, d)\R(s, a, d, s';\vartheta)\bigg) \\
	&\hspace{6em} + \gamma \sum_{a, d, s', \vartheta}\mu(a, \vartheta)\bar\pi^D(d \mid s)\T(s' \mid s, a, d)\big(w(\xi_{a, s'},s')-\xi_{a, s'}(\vartheta)\big)\bigg\}\\
	&= \min_{\bar\pi^D, \xi}\max_{\mu} \bigg\{ \sum_{a, \vartheta}\mu(a,\vartheta) \bigg(\zeta(\vartheta) + \sum_{d, s'}\bar\pi^D(d \mid s)\T(s'\mid s, a, d)\R(s, a, d, s';\vartheta)\bigg) \\
	&\hspace{3em} + \gamma \sum_{a, d, s', \vartheta}\mu(a, \vartheta)\bar\pi^D(d \mid s)\T(s' \mid s, a, d)\big(w'(\xi_{a, s'},s')-\xi_{a, s'}(\vartheta)\big)\bigg\} + \gamma\varepsilon\\
	&=  [Hw'](s, \zeta) + \gamma\varepsilon.
	\end{align*}
	
	Therefore, the one-stage value backup operator is a contraction mapping for a given $(\bar\pi^D, \xi, \mu)$.
\end{proof}

\subsection{Proof of Lemma \ref{lem:sawtooth}}
\begin{proof}
By definition of the subroutine SAWTOOTH-A in Algorithm 1, for given $\Y_s, \W_s$, the  function $\Upsilon_v\big(\Y_s, \W_s, \cdot\big)$ returns the $x_j$ for the $j$ that makes $v_j - c^Tb_j > 0$. Denote this $j$ by $j^*$.  
Let $e_{j,\vartheta}=\big(0,\cdots, 1/b_j(\vartheta),\cdots,0\big)^T\in\Rb^{|\Theta|}$ be  an all-zero vector except that the $\vartheta$-th element is $1/b_j(\vartheta)$. 
 Thus, the following equivalence relationship holds, \emph{i.e.,} for any $V$
\begin{align}\label{equ:x_j_take_min}
V\leq \Upsilon_v\big(\Y_s, \W_s, b\big)\quad   
 \Longleftrightarrow \quad V\leq  c^Tb +\min_{\vartheta\in\Theta}\big\{e_{j^*,\vartheta}^T b| b_{j^*}(\vartheta)>0\big\}\cdot(v_{j^*} - c^Tb_{j^*}). 
\end{align}
The positivity of $v_{j^*} - c^Tb_{j^*}$ implies that \eqref{equ:x_j_take_min} can then equivalently be written as 
\begin{align}\label{equ:x_j_take_min_2}
V\leq \Upsilon_v\big(\Y_s, \W_s, b\big)\quad   
 \Longleftrightarrow \quad V\leq  c^Tb +e_{j^*,\vartheta}^T b \cdot(v_{j^*} - c^Tb_{j^*}), \quad \forall \vartheta\in\Theta, 
\end{align}
which essentially describes $|\Theta|$ constraints that are linear in $b$. Note that the dependences of the constraints \eqref{equ:x_j_take_min_2} on $\W_s$ and $\Y_s$ are implicitly embedded in finding $c$ and $(b_{j^*},v_{j^*})$, respectively. 

Similarly,  SAWTOOTH-D in Algorithm \ref{alg:PBVI_NCIRL} returns the  $y_j$ that makes $w_j - c^T\zeta_j<0$. Denoting this $j$ by $j^*$, we have the following equivalent conditions
\begin{align}\label{equ:y_j_take_max}
V\geq \Upsilon_w\big(\Y_s, \W_s, \zeta\big)\qquad  \Longleftrightarrow \qquad V\geq c^T\zeta +d_{j^*,\vartheta}^T b\cdot(w_j - c^T\zeta_j),\quad \forall \vartheta\in\Theta,
\end{align}
where $d_{j^*,\vartheta}=\big(0,\cdots, 1/\zeta_j(\vartheta),\cdots,0\big)^T\in\Rb^{|\Theta|}$ be  an all-zero vector except that the $\vartheta$-th element is $1/\zeta_j(\vartheta)$.   Note that \eqref{equ:y_j_take_max} describes $|\Theta|$ constraints that are linear in $\zeta$. 
This completes the proof. 
\end{proof}

\clearpage

\section{Details of the NC-PBVI Algorithm}\label{sec:pbvi}

The experimental setup is as follows. We randomly generate attack graphs with sizes ranging from 6 to 10 nodes. Root nodes are assumed to be enabled initially. Furthermore, we limit the in-degree and out-degree of nodes to be at most 3. 
For each graph of size $n$, we run an experiment on a finite horizon of length $n$. The intent parameter is uniformly chosen from a set of random intent parameters of size $|\Theta|=10$. The attacker's accumulated reward is collected at each stage and normalized by the total reward across all nodes. To compare the average performance of N-CIRL and MA-IRL, we run 20 graph instances for each size and plot the attackers' average reward, see Figure \ref{plot:value}. Note that in MA-IRL, both players are playing a complete information game. The difference is that the defender is playing Nash equilibrium strategies based on an intent parameter that is inferred from existing attack data. The attacker, on the other hand, knows its true intent (which is in general different from the defender's inferred intent) and plays its corresponding Nash equilibrium strategies.

All the experiments were run on a machine with an AMD Ryzen 1950X Processor and 32GB of RAM. We used GUROBI 8.1.1 to solve the LPs used in our algorithm and the probability of success, $\beta_{iy}$, is assumed to be 0.8. The detailed pseudocode of the proposed NC-PBVI algorithm is summarized in Algorithm \ref{alg:PBVI_NCIRL}. 

\begin{algorithm}[h!]
	\caption{Non-Cooperative Point-Based Value Iteration (NC-PBVI)}
	\small
	\vspace{0.2em}

	\begin{multicols}{2}
	\hspace{-1em}
		\begin{minipage}[t]{0.49\textwidth}
				\begin{algorithmic}
				\Function{NC-PBVI }{$b_0, \zeta_0, N, T$}
				\For {$p$ in $\{A, D\}$}
				\For {$s \in \S$}
				\State {Initialize $\Y^p_s, \W^p_s$}
				\EndFor
				\For{N expansions}
				\For{T iterations} 
				\State {$\Y^p, \W^p \gets$\Call{ UPDATE-p}{$\Y^p, \W^p$} } 
				\EndFor
				\State{$\Y^p, \W^p \gets$\Call{ EXPAND-p }{$\Y^p, \W^p$}}
				\EndFor 
				\EndFor
				\State {Compute $\pi^A, \pi^D$ by solving $P_A(s, b)$ and}
				\State {$P_D(s, \zeta)$ for all $s$ and finite sets of $b$ (resp. $\zeta$). }
				\EndFunction
		
		    	\vspace{0.5em}

				\Function{UPDATE-A }{$\Y, \W$}
				\For {$s \in \S$}
					\For {$(b, v) \in \Y_s \cup \W_s$}
						\State {update $v$ by solving $P_A(s, b)$}
					\EndFor
				\EndFor
				\State {\Return {$\Y, \W$}}
				\EndFunction
				
				\vspace{0.5em}
				
				\Function{UPDATE-D }{$\Y, \W$}
				\For {$s \in \S$}
					\For {$(\zeta, w) \in \Y_s \cup \W_s$}
						\State {update $w$ by solving $P_D(s, \zeta)$}
					\EndFor
				\EndFor
				\State {\Return {$\Y, \W$}}
				\EndFunction
				
				\vspace{0.5em}
				
				\Function{EXPAND-A }{$\Y, \W$}
				\For {$s \in \S$}
				\For {$(b, v) \in \Y_s\cup \W_s$}
				\State {$\Omega \gets \varnothing$}
				\State {$\bar\pi^A \leftarrow$ solve $P_A(s, b)$}
				\For {$a \in \A(s)$}
				\State{$b_a \gets \tau(s, b, a)$}
				\State {$\Omega \gets \Omega \cup b_a$}
				\EndFor
				\State {$b' \gets \argmax\limits_{b_a \in \Omega} \sum\limits_{\vartheta \in \Theta}|b_{a}(\vartheta) - b(\vartheta)|$}
				\If {$(b', \cdot) \notin \Y_s\cup\W_s$}
				\State {$V_{s, b'} \gets$ solve $P_{A}(s, b')$}
				\State {$\Y_s \gets \Y_s \cup (b', V_{s, b'})$}
				\EndIf
				\EndFor
				\EndFor
				\State {\Return {$\Y, \W$}}
				\EndFunction

			\end{algorithmic}
		\end{minipage}	
	
		\begin{minipage}[t]{0.49\textwidth}
			\begin{algorithmic}
	
				\Function{EXPAND-D }{$\Y, \W$}
				\For {$s \in \S$}
				\For {$(\zeta, w) \in \Y_s \cup \W_s$}
				\State {$\zeta' \leftarrow$ solve $P_D(s, \zeta)$}
				\If {$(\zeta', \cdot) \notin \Y_s\cup\W_s$}
				\State {$W_{s, \zeta'} \gets$ solve $P_{D}(s, \zeta')$}
				\State {$\Y_s \gets \Y_s \cup (\zeta', W_{s, \zeta'})$}
				\EndIf
				\EndFor
				\EndFor
				\State {\Return {$\Y, \W$}}
				\EndFunction
				
				\vspace{0.5em}
				
				\Function{SAWTOOTH-A }{$\Y_s, \W_s, b$}
				\For {$(b_i, v_i) \in \W_s$}
				\State {$c_i \gets v_i$}
				\EndFor
				\State {$x_j \gets c^Tb$}
				\For {$(b_j, v_j) \in \Y_s$}
				\If {$v_j - c^Tb_j > 0$}
				\State {$\phi \gets \min\limits_{\vartheta\in\Theta}\{b(\vartheta)/b_j(\vartheta)| b_j(\vartheta)>0\}$}
				\State {$x_j \gets x_j +\phi(v_j - c^Tb_j)$}
				\State {\textbf{break}}
				\EndIf
				\EndFor
				\State {\Return {~$x_j$}}
				\EndFunction
				
				\vspace{0.5em}
				
				\Function{SAWTOOTH-D }{$\Y_s, \W_s, \zeta$}
				\For {$(\zeta_i, v_i) \in \W_s$}
				\State {$c_i \gets v_i$}
				\EndFor
				\State {$y_j \gets c^T\zeta$}
				\For {$(\zeta_j, w_j) \in \Y_s$}
				\If {$w_j - c^T\zeta_j<0$}
				\State {$\psi \gets \min\limits_{\vartheta\in\Theta}\{\zeta(\vartheta)/\zeta_j(\vartheta) | \zeta_j(\vartheta) >0\}$}
				\State {$y_j \gets y_j +\psi(w_j - c^T\zeta_j)$}
				\State {\textbf{break}}
				\EndIf
				\EndFor
				\State {\Return {~$y_j$}}
				\EndFunction
				
			\end{algorithmic}
		\end{minipage}
	\end{multicols}\label{alg:PBVI_NCIRL}
\end{algorithm}

\end{document}